\def\eqn#1{\eqno(#1)}

\def\mi{\bigskip\noindent}

\documentclass[russian,12pt]{article}
\usepackage[T1]{fontenc}
\usepackage[cp866]{inputenc}
\usepackage{babel}
\usepackage{psfig}
\begin{document}
{\center

\mi{\bf О ЛИНЕЙНОЙ УСТОЙЧИВОСТИ СТАЦИОНАРНЫХ ПРОСТРАНСТВЕННО-ПЕРИОДИЧЕСКИХ\\
МАГНИТОГИДРОДИНАМИЧЕСКИХ СИСТЕМ\\ К ДЛИННОПЕРИОДНЫМ ВОЗМУЩЕНИЯМ}

\mi В.А.Желиговский

\mi Международный институт теории прогноза землетрясений\\
и математической геофизики РАН

\mi Лаборатория общей аэродинамики, Институт механики МГУ;\\

\mi Обсерватория Лазурного Берега, CNRS U.M.R. 6529\\

\mi{\it E-mail}: vlad@mitp.ru

}

\bigskip Построено полное асимптотическое разложение решений задачи о линейной
устойчивости трехмерных стационарных пространст\-венно-периодических
МГД состояний по отношению к длиннопериодным возмущениям.
Найден тензор вих\-ревой диффузии для стационарных состояний, обладающих
центральной симмет\-рией. Численно показано, что эффект отрицательной вихревой
диффузии появля\-ется при учете наличия возмущений поля скорости при существенно
б\'ольших значениях коэффициента молекулярной магнитной диффузии, чем в задаче
кине\-матического динамо (в которой предполагается отсутствие возмущений поля
скорости).

\mi{\bf 1. Введение.}

Согласно общепринятым представлениям магнитное поле Земли обязано
сво\-им существованием магнитоконвекционным процессам во внешнем ядре.
В прин\-ципе эти процессы можно численно моделировать, решая соответствующую
сис\-тему трехмерных уравнений -- уравнение Навье-Стокса с учетом сил Лоренца и
термоконвекционной плавучести, уравнение магнитной индукции и уравнение
теплопроводности. Этот подход применен, например, в [Glatzmaier, Roberts,
1995, 1996; Roberts, Glatzmaier, 2001; Sarson, Jones, 1999]
для изучения инверсий магнит\-ного поля Земли.
Однако он имеет внутренние ограничения, связанные c недоста\-точной
вычислительной мощностью существующих компьютеров -- разрешение, которое можно
получить в расчетах, мал\'о для адекватного представления реше\-ний при
значениях определяющих констант, соответствующих геофизическим условиям.
Таким образом, определенную ценность представляют аналитические решения
идеализированных задач.

Одна из таких задач рассмотрена в данной работе -- задача об устойчивости
магнитогидродинамических стационарных состояний. Предполагается, что харак\-терный
пространственный масштаб рассматриваемого стационарного состояния существенно
меньше характерного масштаба возмущений. Отношение этих масш\-табов, $\epsilon$,
можно тогда рассматривать как малый параметр, что позволяет приме\-нить
асимптотические методы и построить решения задачи об устойчивости в виде
степенных рядов по этому параметру. Показано, что для систем общего положения
малые возмущения подвержены действию $\gamma$-эффекта. Как следствие, такие
системы неустойчивы: главные члены разложений собственных значений,
соответствующих противоположным волновым векторам, отвечающим длинным масштабам
возмущений, имеют противоположные знаки.

В случае, если система обладает симметрией относительно центра (и тем самым
не является системой общего положения), $\gamma$-эффект отсутствует,
а главные члены разложений собственных значений и собственных векторов
являются соот\-ветственно
собственными значениями и собственными векторами так называемого оператора
анизотропной комбинированной вихревой (турбулентной) диффузии. Он является
оператором в частных производных второго порядка и задается уравнениями (26),
коэффициенты (27) которого требуют решения т.н. вспомога\-тельных краевых задач
(11), (12), (24) и (25) (см. ниже). Данные во вспомогатель\-ных задачах
имеют единственный характерный пространственный масштаб -- тот же самый,
что и рассматриваемое стационарное состояние -- и поэтому допускают
численное решение. Оператор вихревой диффузии, в отличие от оператора
моле\-кулярной диффузии, может не быть знакоопределен. В случае, если он имеет
положительные собственные значения, говорят о возникновении феномена
отри\-цательной диффузии. Проведенные
численные эксперименты показывают, что вихревая диффузия может принимать
отрицательные значения при существенно б\'ольших значениях коэффициента
молекулярной магнитной диффузии в рас\-сматриваемой ситуации (т.е. при учете
наличия возмущений поля скорости), чем в ситуации кинематического динамо.

Отметим, что указанные механизмы возникновения неустойчивости
магнито\-гидродинамического стационарного состояния являются слабыми:
при наличии $\gamma$-эффекта главный член разложения собственных значений
пропорционален $\epsilon$, а для систем с центром симметрии пропорционален
$\epsilon^2$.

Рассматриваемое стационарное магнитогидродинамическое состояние
${\bf V},{\bf H},P$ удовлетворяет системе уравнений
$$\nu\Delta{\bf V}+{\bf V}\times(\nabla\times{\bf V})
+(\nabla\times{\bf H})\times{\bf H}-\nabla P+{\bf F}=0,\eqn{1.1}$$
$$\eta\Delta{\bf H}+\nabla\times({\bf V}\times{\bf H})+{\bf J}=0,\eqn{1.2}$$
$$\nabla\cdot{\bf V}=\nabla\cdot{\bf H}=0.\eqn{1.3}$$
Здесь ${\bf V}({\bf x})$ -- скорость потока проводящей жидкости,
${\bf H}({\bf x})$ -- магнитное поле, $P({\bf x})$ -- давление, $\nu$ и $\eta$ --
коэффициенты кинематической и магнитной молекуляр\-ной диффузии, соответственно,
${\bf F}({\bf x})$ -- объемная сила, ${\bf J}({\bf x})$
отвечает наличию в системе распределения наложенных внешних токов.
Поля $2\pi$-периодичны по пространственным переменным, их средние
по пространству равны 0. Исследова\-ние линейной устойчивости этого
стационарного состояния приводит к задаче на собственные значения
$$\nu\Delta{\bf v}+{\bf v}\times(\nabla\times{\bf V})
+{\bf V}\times(\nabla\times{\bf v})+(\nabla\times{\bf h})\times{\bf H}
+(\nabla\times{\bf H})\times{\bf h}-\nabla p=\lambda{\bf v},\eqn{2.1}$$
$$\eta\Delta{\bf h}+\nabla\times({\bf v}\times{\bf H})
+\nabla\times({\bf V}\times{\bf h})=\lambda{\bf h},\eqn{2.2}$$
$$\nabla\cdot{\bf v}=\nabla\cdot{\bf h}=0.\eqn{2.3}$$
(Уравнения (2) получены стандартной процедурой отбрасывания квадратичных
членов в полных уравнениях, описывающих эволюцию возмущений,
см., например, [Chandrasekhar, 1961].)
Предполагается, что профили возмущений ${\bf v}({\bf x}),{\bf h}({\bf x})$, $p({\bf x})$
имеют прост\-ранственный масштаб $2\pi/\epsilon$, где $\epsilon>0$
-- малый параметр. В данной работе построено полное асимптотическое разложение
для профилей возмущений и инкрементов их роста (или убывания) во времени
в виде степенных рядов по $\epsilon$ (при выполнении определенных условий,
описанных ниже).

При рассмотрении указанных задач используется математическая теория\break осреднения
дифференциальных операторов [Жиков, Козлов, Олейник, Ха Тьен Нгоан, 1979; Козлов,
1978; Bensoussan, Lions, Papanicolaou, 1978]. Из аналогичных представленным
здесь разложений для немагнитных систем \hbox{(${\bf h}={\bf H}=0$)} следует,
что эффект отрицательной вихревой диффузии может возникнуть в двумерных
[Sivashinsky, Yakhot, 1985; Sivashinsky, Frenkel, 1992;
Gama, Vergassola, Frisch, 1994] и трехмерных [Dubrulle, Frisch, 1991;
Wirth, Gama, Frisch, 1995] гидродина\-мических системах, если поле
скорости обладает центральной симметрией или является полем Бельтрами.
В гидродинамических системах, не имеющих этих свойств,
подобные разложения указывают на наличие т.н. АКА-эффекта (кинема\-тического
$\alpha$-эффекта) [Frisch, Zhen Su She, Sulem, 1987; Sulem, She, Scholl,
Frisch, 1989]. При переносе пассивного скаляра вихревая диффузия только
усиливает молекулярную диффузию [Biferale, Crisanti, Vergassola, Vulpiani,
1995; Vergassola, Avellaneda, 1997]. В задаче кинематического динамо
(об устойчивости магнит\-ного поля при фиксированном поле скорости $\bf V$
проводящей среды) асимптотичес\-кие разложения предсказывают
появление $\alpha$- [Вишик, 1986; Желиговский, 1990, 1991;
Zheligovsky, 1991] или $\gamma$-эффекта [Вишик, 1987], а если у поля
скорости есть центр симметрии -- возможность генерации магнитного поля
посредством механизма отрицательной вихревой диффузии [Lanotte, Noullez,
Vergassola, Wirth, 1999; Zheligovsky, Podvigina, Frisch, 2001].

\mi{\bf 2. Формальные асимптотические разложения.}

1. Пусть $\bf x$, как и выше, обозначает быструю пространственную переменную,
а ${\bf y}=\epsilon\bf x$ -- медленную.
Решение задачи на собственные значения (2) ищем в виде степенных рядов
$${\bf v}=\sum_{n=0}^\infty({\bf v}_n({\bf y})+{\bf w}_n({\bf x},{\bf y}))
\epsilon^n,\quad
{\bf h}=\sum_{n=0}^\infty({\bf h}_n({\bf y})+{\bf g}_n({\bf x},{\bf y}))
\epsilon^n,\eqn{3}$$
$$p=\sum_{n=0}^\infty(p_n({\bf y})+q_n({\bf x},{\bf y}))\epsilon^n,\eqn{4}$$
$$\lambda=\sum_{n=0}^\infty\lambda_n\epsilon^n.\eqn{5}$$
В (3),(4) ${\bf v}_n,{\bf h}_n$ и $p_n$ -- средние части, а ${\bf w}_n,{\bf g}_n$
и $q_n$ -- соответственно, осциллирующие части соответствующих членов
разложений. Имеется в виду усреднение по быст\-рым переменным:
$$\langle{\bf f}\rangle\equiv(2\pi)^{-3}\int_{T^3}{\bf f}({\bf x},{\bf y})d{\bf x}$$
-- средняя, а $\{{\bf f}\}\equiv{\bf f}-\langle{\bf f}\rangle$ -- осциллирующая
часть поля ${\bf f}({\bf x},{\bf y})$.

Приравняем нулю коэффициенты рядов по степеням $\epsilon$, полученных
подстанов\-кой рядов (3) в (2.3).
Выделяя среднюю и осциллирующую часть полученных уравнений, находим
$$\nabla_{\bf y}\cdot{\bf v}_n=\nabla_{\bf y}\cdot{\bf h}_n=0,\eqn{6.1}$$
$$\nabla_{\bf x}\cdot{\bf w}_n+\nabla_{\bf y}\cdot{\bf w}_{n-1}=
\nabla_{\bf x}\cdot{\bf g}_n+\nabla_{\bf y}\cdot{\bf g}_{n-1}=0\eqn{6.2}$$
для всех $n\ge0$. Здесь в дифференциальных операторах с индексами $\bf x$ и
$\bf y$ дифференцирование производится по быстрым и медленным пространственным
переменным, соответственно (все члены разложений с индексом $n<0$
по опреде\-лению равны 0).

Введем обозначения
$$L^v({\bf w},{\bf g},q)\equiv\nu\Delta_{\bf x}{\bf w}
+{\bf V}\times(\nabla_{\bf x}\times{\bf w})+{\bf w}\times(\nabla_{\bf x}\times{\bf V})$$
$$+(\nabla_{\bf x}\times{\bf H})\times{\bf g}
+(\nabla_{\bf x}\times{\bf g})\times{\bf H}-\nabla_{\bf x}q,$$
$$L^h({\bf w},{\bf g})\equiv\eta\Delta_{\bf x}{\bf g}
+\nabla_{\bf x}\times({\bf V}\times{\bf g}+{\bf w}\times{\bf H}).$$
Отметим, что в силу (1.3) и $2\pi$-периодичности $\bf V$ и $\bf H$
$$\langle L^v({\bf w},{\bf g},q)\rangle=\langle{\bf V}\nabla_{\bf x}
\cdot{\bf w}-{\bf H}\nabla_{\bf x}\cdot{\bf g}\rangle,\qquad
\langle L^h({\bf w},{\bf g})\rangle=0\eqn{7}$$
для произвольных полей ${\bf w}({\bf x},{\bf y}),{\bf g}({\bf x},{\bf y}),
q({\bf x},{\bf y})$, $2\pi$-периодических по быст\-рым переменным.

В дальнейшем предполагаем, что для произвольных
${\bf f}^v({\bf x}),{\bf f}^h({\bf x})$ из простран\-ства соленоидальных
полей с нулевым средним, $2\pi$-периодических по быстрым переменным,
система уравнений
$$L^v({\bf w},{\bf g},q)={\bf f}^v,\quad L^h({\bf w},{\bf g})={\bf f}^h,\quad
\nabla_{\bf x}\cdot{\bf w}=\nabla_{\bf x}\cdot{\bf g}=0\eqn{8}$$
имеет единственное решение ${\bf w},{\bf g},q$ в указанном пространстве.
Это условие выпол\-нено для стационарных состояний (1) общего положения
(т.е. малое возмущение полей $\bf F$ и $\bf J$ в системах, где оно не выполнено,
приводит к его выполнению).

Подставляя (3)-(5) в уравнения (2.1) и (2.2), преобразуем последние
к виду равенств рядов по степеням $\epsilon$. Приравнивая коэффициенты
этих рядов, получаем рекуррентную систему уравнений, которую последовательно
решаем совместно с условиями (6),
выделяя среднюю и осциллирующую часть каждого уравнения.

2. Из главных членов (порядка $\epsilon^0$) рядов (2.1) и (2.2)
получаем уравнения
$$L^v({\bf w}_0,{\bf g}_0,q_0)+{\bf v}_0\times(\nabla_{\bf x}\times{\bf V})
+(\nabla_{\bf x}\times{\bf H})\times{\bf h}_0=\lambda_0({\bf v}_0+{\bf w}_0),\eqn{9.1}$$
$$L^h({\bf w}_0,{\bf g}_0)
+({\bf h}_0\cdot\nabla_{\bf x}){\bf V}-({\bf v}_0\cdot\nabla_{\bf x}){\bf H}
=\lambda_0({\bf h}_0+{\bf g}_0).\eqn{9.2}$$

Усредняя их с использованием (7), получаем в силу равенств
(6.2) при $n=0$, что $0=\lambda_0{\bf v}_0=\lambda_0{\bf h}_0$,
откуда $\lambda_0=0$ (альтернативная возможность ${\bf v}_0={\bf h}_0=0$ не
представляет интереса, т.к. она соответствует возмущению короткопериодной моды).

Осциллирующие части уравнений (9) в силу их линейности имеют
тогда решения следующей структуры:
$${\bf w}_0=\sum_{k=1}^3({\bf S}^{v,v}_k{\bf v}^k_0+{\bf S}^{h,v}_k{\bf h}^k_0),\qquad
{\bf g}_0=\sum_{k=1}^3({\bf S}^{v,h}_k{\bf v}^k_0+{\bf S}^{h,h}_k{\bf h}^k_0),\eqn{10.1}$$
$$q_0=\sum_{k=1}^3(S^{v,p}_k{\bf v}^k_0+S^{h,p}_k{\bf h}^k_0),\eqn{10.2}$$
где $2\pi$-периодические функции $\bf S$ с нулевым средним являются решениями
систем уравнений
$$L^v({\bf S}^{v,v}_k,{\bf S}^{v,h}_k,S^{v,p}_k)=
-{\bf e}_k\times(\nabla_{\bf x}\times{\bf V}),\eqn{11.1}$$
$$\nabla_{\bf x}\cdot{\bf S}^{v,v}_k=0,\eqn{11.2}$$
$$L^h({\bf S}^{v,v}_k,{\bf S}^{v,h}_k)=
{\partial{\bf H}\over\partial{\bf x}_k};\eqn{11.3}$$
$$L^v({\bf S}^{h,v}_k,{\bf S}^{h,h}_k,S^{h,p}_k)=
{\bf e}_k\times(\nabla_{\bf x}\times{\bf H}),\eqn{12.1}$$
$$\nabla_{\bf x}\cdot{\bf S}^{h,v}_k=0,\eqn{12.2}$$
$$L^h({\bf S}^{h,v}_k,{\bf S}^{h,h}_k)=
-{\partial{\bf V}\over\partial{\bf x}_k}.\eqn{12.3}$$
Здесь ${\bf e}_k$ -- единичный вектор вдоль оси координат $x_k$, верхний
индекс $k$ нумерует компоненты вектора:
$${\bf v}_0=\sum_{k=1}^3{\bf v}^k_0{\bf e}_k,\quad
{\bf h}_0=\sum_{k=1}^3{\bf h}^k_0{\bf e}_k.$$
Задачи (11) и (12) имеют единственное решение согласно
изначальному предполо\-жению о разрешимости задач (8).
Взяв дивергенцию от (11.3) и (12.3), находим
$\nabla_{\bf x}\cdot{\bf S}^{v,h}_k=\nabla_{\bf x}\cdot{\bf S}^{h,h}_k=0$.

3. Рассмотрим уравнения, полученные из членов (2.1) и (2.2) порядка
$\epsilon^1$. Используя (7), (6.2) при $n=1$ и
(10.1), их средние части представим в виде
$$\sum_{k=1}^3(\Gamma^{v,v}_k\nabla_{\bf y}{\bf v}^k_0
+\Gamma^{h,v}_k\nabla_{\bf y}{\bf h}^k_0)-\nabla_{\bf y}p'_0=
\lambda_1{\bf v}_0,\eqn{13.1}$$
$$\nabla_{\bf y}\times\sum_{k=1}^3(\Gamma^{v,h}_k{\bf v}^k_0+
\Gamma^{h,h}_k{\bf h}^k_0)=\lambda_1{\bf h}_0,\eqn{13.2}$$
где $p'_0\equiv p_0-\langle{\bf V}\cdot{\bf w}_0-{\bf H}\cdot{\bf g}_0\rangle$,
элементы матриц $\Gamma^{\cdot,v}_k$ размера $3\times3$ имеют вид
$$(\Gamma^{v,v}_k)^m_j\equiv\langle
-{\bf V}^m({\bf S}^{v,v}_k)^j-{\bf V}^j({\bf S}^{v,v}_k)^m
+{\bf H}^m({\bf S}^{v,h}_k)^j+{\bf H}^j({\bf S}^{v,h}_k)^m\rangle,$$
$$(\Gamma^{h,v}_k)^m_j\equiv\langle
-{\bf V}^m({\bf S}^{h,v}_k)^j-{\bf V}^j({\bf S}^{h,v}_k)^m
+{\bf H}^m({\bf S}^{h,h}_k)^j+{\bf H}^j({\bf S}^{h,h}_k)^m\rangle,$$
а трехмерные векторы $\Gamma^{\cdot,h}_k$ задаются выражениями
$$\Gamma^{v,h}_k\equiv\langle
{\bf V}\times{\bf S}^{v,h}_k-{\bf H}\times{\bf S}^{v,v}_k\rangle,\quad
\Gamma^{h,h}_k\equiv\langle
{\bf V}\times{\bf S}^{h,h}_k-{\bf H}\times{\bf S}^{h,v}_k\rangle.$$

Дальнейший ход решения уравнений, полученных из (2.1) и (2.2) при $n>0$,
для случаев $\gamma$-эффекта и вихревой диффузии различен.

\mi{\bf 3. $\gamma$-эффект.}

4. Собственные функции задачи на собственные значения (13) являются
гармо\-никами Фурье:
${\bf v}_0=\hat{\bf v}_0e^{i\xi\bf y},\ {\bf h}_0=\hat{\bf h}_0e^{i\xi\bf y}$,
где $\xi$ -- произвольный постоянный волновой вектор, а
векторы $\hat{\bf v}_0$ и $\hat{\bf h}_0$ удовлетворяют следующим условиям:
$$-{i\over|\xi|^2}\xi\times\left(\xi\times\sum_{k=1}^3(\hat{\bf v}^k_0\Gamma^{v,v}_k
+\hat{\bf h}^k_0\Gamma^{h,v}_k)\xi\right)=\lambda_1\hat{\bf v}_0,\eqn{14.1}$$
$$i\xi\times\sum_{k=1}^3(\hat{\bf v}^k_0\Gamma^{v,h}_k+
\hat{\bf h}^k_0\Gamma^{h,h}_k)=\lambda_1\hat{\bf h}_0.\eqn{14.2}$$
Условия соленоидальности (6.1) преобразуются к виду
$$\hat{\bf v}_0\cdot\xi=\hat{\bf h}_0\cdot\xi=0.\eqn{15}$$

Обозначим матрицу размера $6\times6$ в левой части (14), которая умножается на
вектор $(\hat{\bf v}_0,\hat{\bf h}_0)$, через $\Gamma(\xi)$. Отображение
$(\hat{\bf v}_0,\hat{\bf h}_0)\to\Gamma(\xi)(\hat{\bf v}_0,\hat{\bf h}_0)$
является символом предельного псевдодифференциального оператора
$L_\infty$, действующего на $({\bf v}_0({\bf y}),{\bf h}_0({\bf y}))$,
который определен левой частью (13). Пусть
$$V_\xi=\{(\hat{\bf v},\hat{\bf h})\ |\ \hat{\bf v},\hat{\bf h}\in R^3,
\hat{\bf v}\cdot\xi=\hat{\bf h}\cdot\xi=0\}\subset R^6,$$
$$\bar{H}^s=\{{\bf f}({\bf y})\in H^s(T^3)\ |\ \nabla\cdot{\bf f}=0\},$$
где $H^s(T^3)$ -- пространство Соболева c метрикой $W^s_2(T^3)$ соленоидальных
функций с нулевым средним на трехмерном торе. Отметим следующее свойство
спектра $L_\infty$: если $\lambda$ -- собственное значение, отвечающее
некоторому волновому вектору $\xi$, то $-\lambda$ -- собственное значение,
отвечающее $-\xi$. Таким образом, имеет место характерная для
$\gamma$-эффекта альтернатива: либо возмущение с любым волновым вектором
имеет вид гармонических колебаний с постоянной амплитудой, либо
существуют волновые векторы, для которых возмущения неустойчивы.

5. Дальнейшее изложение в данном разделе проведем для случая, когда для
предельного оператора выполнено следующее условие типа эллиптичности:
для любого $\xi$ матрица $\Gamma(\xi)$ определяет линейный оператор,
действующий в $V_\xi$ невы\-рожденно. Тогда образ линейного оператора
$L_\infty:\bar{H}^{s+1}\to\bar{H}^s$ замкнут, $L_\infty$
имеет дискретный спектр, и для любого $\lambda$, принадлежащего
резольвентному множеству $L_\infty$, оператор $(L_\infty-\lambda)^{-1}$
компактен.

Кроме того, будем предполагать для простоты, что кратность собственного значения
$\lambda_1$ матрицы $\Gamma(\xi)$ равна 1. Это последнее условие не вызвано
существом дела -- аналогичные (несколько более громоздкие) построения можно
провести и в случае кратного собственного значения. Оно выполнено для систем
общего положения, т.е. если для некоторого стационарного состояния
рассматриваемое собственное значение оказалось кратным, то при почти любом
малом возмущении функций $\bf F$ и $\bf J$, определяющих стационарное
состояние системы (1), это собствен\-ное значение расщепляется на соответствующее
число собственных значений кратности 1.

Для решения уравнений, полученных из членов (2.1) и (2.2)
порядка $\epsilon^n$ при $n>0$, удобно сделать подстановки
$${\bf w}_n={\bf w}'_n+\sum_{k=1}^3({\bf S}_k^{v,v}{\bf v}^k_n+{\bf S}^{h,v}_k{\bf h}^k_n),\eqn{16.1}$$
$${\bf g}_n={\bf g}'_n+\sum_{k=1}^3({\bf S}_k^{v,h}{\bf v}^k_n+{\bf S}^{h,h}_k{\bf h}^k_n),\eqn{16.2}$$
$$q_n=q'_n+\sum_{k=1}^3(S^{v,p}_k{\bf v}^k_n+S^{h,p}_k{\bf h}^k_n),\eqn{16.3}$$
после чего эти уравнения принимают вид
$$L^v({\bf w}'_n,{\bf g}'_n,q'_n)
+\nu\left(2(\nabla_{\bf x}\cdot\nabla_{\bf y}){\bf w}_{n-1}
+\Delta_{\bf y}({\bf v}_{n-2}+{\bf w}_{n-2})\right)$$
$$+{\bf V}\times(\nabla_{\bf y}\times({\bf v}_{n-1}+{\bf w}_{n-1}))
+(\nabla_{\bf y}\times({\bf h}_{n-1}+{\bf g}_{n-1}))\times{\bf H}$$
$$-\nabla_{\bf y}(p_{n-1}+q_{n-1})
-\sum_{m=0}^{n-1}\lambda_{n-m}({\bf v}_m+{\bf w}_m)=0,\eqn{17.1}$$
$$L^h({\bf w}'_n,{\bf g}'_n)
+\eta\left(2(\nabla_{\bf x}\cdot\nabla_{\bf y}){\bf g}_{n-1}
+\Delta_{\bf y}({\bf h}_{n-2}+{\bf g}_{n-2})\right)$$
$$-({\bf V}\cdot\nabla_{\bf y}){\bf h}_{n-1}+({\bf H}\cdot\nabla_{\bf y}){\bf v}_{n-1}$$
$$+\nabla_{\bf y}\times({\bf V}\times{\bf g}_{n-1}+{\bf w}_{n-1}\times{\bf H})
-\sum_{m=0}^{n-1}\lambda_{n-m}({\bf h}_m+{\bf g}_m)=0.\eqn{17.2}$$
Покажем, что (17) можно последовательно решить, причем
$${\bf v}_n({\bf y})=\hat{\bf v}_ne^{i\xi\bf y},\quad
{\bf h}_n({\bf y})=\hat{\bf h}_ne^{i\xi\bf y},\quad
p_n({\bf y})=\hat{p}_ne^{i\xi\bf y},\eqn{18.1}$$
$${\bf w}_n({\bf x},{\bf y})=\hat{\bf w}_n({\bf x})e^{i\xi\bf y},\quad
{\bf g}_n({\bf x},{\bf y})=\hat{\bf g}_n({\bf x})e^{i\xi\bf y},\quad
q_n({\bf x},{\bf y})=\hat{q}_n({\bf x})e^{i\xi\bf y}.\eqn{18.2}$$
Предположим, что для некоторого $N$ решены все уравнения при $n<N$
и опреде\-лены неизвестные функции ${\bf v}_n,{\bf w}_n,{\bf h}_n,{\bf g}_n,p_n,q_n$
при всех $n<N-1$, ${\bf w}'_{N-1},{\bf g}'_{N-1}$, $q'_{N-1}$, а также $\lambda_n$
при всех $n<N$, причем зависимость неизвестных функций от медленной переменной
выражается согласно (18) в их пропорциональности $e^{i\xi\bf y}$.
Рассмотрим уравнение, соответствующее $n=N$.

Подстановка (16) при $n=N-1$ приводит средние части уравнений (17) при $n=N$
к виду
$$\sum_{k=1}^3(\Gamma^{v,v}_k\nabla_{\bf y}{\bf v}^k_{N-1}
+\Gamma^{h,v}_k\nabla_{\bf y}{\bf h}^k_{N-1})-\nabla_{\bf y}p'_{N-1}
-\lambda_1{\bf v}_{N-1}-\lambda_N{\bf v}_0$$
$$=-\langle{\bf V}\times(\nabla_{\bf y}\times{\bf w}'_{N-1})
-{\bf V}\nabla_{\bf y}\cdot{\bf w}'_{N-1}
+(\nabla_{\bf y}\times{\bf g}'_{N-1})\times{\bf H}
+{\bf H}\nabla_{\bf y}\cdot{\bf g}'_{N-1}\rangle$$
$$-\nu\Delta_{\bf y}{\bf v}_{N-2}+\sum_{m=1}^{N-2}\lambda_{N-m}{\bf v}_m,\eqn{19.1}$$
где
$$p'_{N-1}=p_{N-1}-\left\langle{\bf V}\cdot\sum_{k=1}^3
({\bf S}_k^{v,v}{\bf v}^k_{N-1}+{\bf S}^{h,v}_k{\bf h}^k_{N-1})\right.$$
$$\left.-{\bf H}\cdot\sum_{k=1}^3({\bf S}_k^{v,h}{\bf v}^k_{N-1}
+{\bf S}^{h,h}_k{\bf h}^k_{N-1})\right\rangle,$$
и
$$\nabla_{\bf y}\times\sum_{k=1}^3(\Gamma^{v,h}_k{\bf v}^k_{N-1}+
\Gamma^{h,h}_k{\bf h}^k_{N-1})-\lambda_1{\bf h}_{N-1}-\lambda_N{\bf h}_0$$
$$=-\eta\Delta_{\bf y}{\bf h}_{N-2}-\nabla_{\bf y}\times\langle
{\bf V}\times{\bf g}'_{N-1}+{\bf w}'_{N-1}\times{\bf H}\rangle
+\sum_{m=1}^{N-2}\lambda_{N-m}{\bf h}_m.\eqn{19.2}$$
Согласно предположению индукции правые части (19) имеют вид
$\hat{\bf f}^v_Ne^{i\xi\bf y}$ и $\hat{\bf f}^h_Ne^{i\xi\bf y}$, где
векторы-константы $\hat{\bf f}^v_N$ и $\hat{\bf f}^h_N$ уже известны, причем
$\hat{\bf f}^h_N\cdot\xi=0$. Используя (18.1) при $n=N-1$, из (19) получаем
$$(\Gamma(\xi)-\lambda_1)\left(\hat{\bf v}_{N-1}\atop\hat{\bf h}_{N-1}\right)
-\lambda_N\left(\hat{\bf v}_0\atop\hat{\bf h}_0\right)
=\left(-|\xi|^{-2}\xi\times(\xi\times\hat{\bf f}^v_N)
\atop\hat{\bf f}^h_N\right).\eqn{20}$$
Эту задачу рассматриваем в $V_\xi$. Проектируя (20) на $(\hat{\bf v}_0,\hat{\bf h}_0)\in V_\xi$,
находим $\lambda_N$. Считаем, что компонента вектора
$(\hat{\bf v}_{N-1},\hat{\bf h}_{N-1})$ из подпростран\-ства, натянутого на
$(\hat{\bf v}_0,\hat{\bf h}_0)$, равна нулю (это условие нормировки соответствует тому
факту, что собственную функцию можно умножить на произвольную аналитическую
функ\-цию~$\epsilon$). Поскольку в силу исходного предположения матрица
$\Gamma(\xi)-\lambda_1$ обратима в дополнительном собственном подпространстве
прост\-ранства $V_\xi$, из (20) находим $(\hat{\bf v}_{N-1},\hat{\bf h}_{N-1})$.
Наконец, (19.1) определяет $p'_{N-1}$,
а (16) при $n=N-1$ -- функции ${\bf w}_{N-1},{\bf g}_{N-1}$ и $q_{N-1}$.

Осциллирующие части уравнений (17) при $n=N$, рассматриваемые при условиях
$\nabla_{\bf x}\cdot{\bf w}'_N=-\nabla_{\bf y}\cdot{\bf w}_{N-1}$ и
\hbox{$\nabla_{\bf x}\cdot{\bf g}'_N=-\nabla_{\bf y}\cdot{\bf g}_{N-1}$}
(следствиях (6.2)~), можно тогда решить, выделяя градиентную часть неизвестных
вектор-функций. Пусть $\varphi^v_N$ и $\varphi^h_N$
-- $2\pi$-периодические по быстрой переменной решения краевых задач
$\Delta_{\bf x}\varphi^v_N=-\nabla_{\bf y}\cdot{\bf w}_{N-1}$ и
$\Delta_{\bf x}\varphi^h_N=-\nabla_{\bf y}\cdot{\bf g}_{N-1}$.
Подстановки\break${\bf w}'_N={\bf w}''_N+\nabla_{\bf x}\varphi^v_N$ и
${\bf g}'_N={\bf g}''_N+\nabla_{\bf x}\varphi^h_N$
приводят осциллирующие части уравнений (17) к задаче относительно
${\bf w}''_N,{\bf g}''_N$
вида (8) с условием соленоидальности неизве\-стных векторных полей.

Таким образом, построено полное асимптотическое разложение мод линейных
возмущений и их временн\'ых инкрементов для случая, когда для предельного
оператора $L_\infty$ выполнено приведенное выше условие типа
эллиптичности. Точнее, можно показать, что в этом случае каждому
собственному значению оператора $L_\infty$ отвечает ветвь собственных
значений исходного оператора (задачи (2)~), причем асимптотические ряды (3)-(5)
сходятся при достаточно малых $\epsilon$. Доказа\-тельство этого основано
на применении теории возмущения операторов [Като, 1972] и следует
доказательству [Вишик, 1987] для чисто магнитного случая.

\pagebreak
\mi{\bf 4. Вихревая диффузия.}

6. В данном разделе рассмотрен случай, когда стационарные поля имеют центр симметрии,
который без потери общности считаем расположенным в начале координат:
$${\bf V}({\bf x})=-{\bf V}(-{\bf x}),\quad{\bf H}({\bf x})=-{\bf H}(-{\bf x}).\eqn{21}$$
Тогда область определения оператора $(L^v,L^h)$ распадается на
два собственных подпространства, состоящие из
симметричных (${\bf f}({\bf x})={\bf f}(-{\bf x})$~) и антисимметрич\-ных
(${\bf f}({\bf x})=-{\bf f}(-{\bf x})$~) полей. Поэтому выполнены симметрии
$${\bf S}^{v,v}_k({\bf x})={\bf S}^{v,v}_k(-{\bf x}),\quad
{\bf S}^{v,h}_k({\bf x})={\bf S}^{v,h}_k(-{\bf x}),\quad
S^{v,p}_k({\bf x})=-S^{v,p}_k(-{\bf x}),\eqn{22.1}$$
$${\bf S}^{h,v}_k({\bf x})={\bf S}^{h,v}_k(-{\bf x}),\quad
{\bf S}^{h,h}_k({\bf x})={\bf S}^{h,h}_k(-{\bf x}),\quad
S^{h,p}_k({\bf x})=-S^{h,p}_k(-{\bf x}),\eqn{22.2}$$
и следовательно $\Gamma(\xi)$ -- нулевая матрица.
Уравнения (13) сводятся тогда к\break$-\nabla_{\bf y}p_0=\lambda_1{\bf v}_0$
и $0=\lambda_1{\bf h}_0$, откуда $\lambda_1=0$ и $p_0=0$.

7. Осциллирующие части уравнений (17) при $n=1$
$$L^v({\bf w}'_1,{\bf g}'_1,q'_1)=
-2\nu(\nabla_{\bf x}\cdot\nabla_{\bf y}){\bf w}_0
-{\bf V}\times(\nabla_{\bf y}\times({\bf v}_0+{\bf w}_0))$$
$$-(\nabla_{\bf y}\times({\bf h}_0+{\bf g}_0))\times{\bf H}+\nabla_{\bf y}q_0,$$
$$L^h({\bf w}'_1,{\bf g}'_1)=-2\eta(\nabla_{\bf x}\cdot\nabla_{\bf y}){\bf g}_0$$
$$+({\bf V}\cdot\nabla_{\bf y})({\bf h}_0+{\bf g}_0)
-({\bf H}\cdot\nabla_{\bf y})({\bf v}_0+{\bf w}_0)
-{\bf V}\nabla_{\bf y}\cdot{\bf g}_0+{\bf H}\nabla_{\bf y}\cdot{\bf w}_0,$$
в силу (10) и линейности этих уравнений имеют решения следующей структуры:
$${\bf w}'_1=\sum_{k=1}^3\sum_{m=1}^3({\bf G}^{v,v}_{m,k}{\partial{\bf v}^k_0\over\partial{\bf y}_m}
+{\bf G}^{h,v}_{m,k}{\partial{\bf h}^k_0\over\partial{\bf y}_m}),\eqn{23.1}$$
$${\bf g}'_1=\sum_{k=1}^3\sum_{m=1}^3({\bf G}^{v,h}_{m,k}{\partial{\bf v}^k_0\over\partial{\bf y}_m}
+{\bf G}^{h,h}_{m,k}{\partial{\bf h}^k_0\over\partial{\bf y}_m}),\eqn{23.2}$$
$$q'_1=\sum_{k=1}^3\sum_{m=1}^3(G^{v,p}_{m,k}{\partial{\bf v}^k_0\over\partial{\bf y}_m}
+G^{h,p}_{m,k}{\partial{\bf h}^k_0\over\partial{\bf y}_m}),\eqn{23.3}$$
где функции $\bf G$ являются решениями систем уравнений
$$L^v({\bf G}^{v,v}_{m,k},{\bf G}^{v,h}_{m,k},G^{v,p}_{m,k})=
-2\nu{\partial{\bf S}^{v,v}_k\over\partial{\bf x}_m}-{\bf V}^k{\bf e}_m
+{\bf V}^m{\bf e}_k-({\bf V}\cdot{\bf S}^{v,v}_k){\bf e}_m$$
$$+{\bf V}^m{\bf S}^{v,v}_k+{\bf e}_mS^{v,p}_k
+({\bf H}\cdot{\bf S}^{v,h}_k){\bf e}_m-{\bf H}^m{\bf S}^{v,h}_k\eqn{24.1}$$
$$\nabla_{\bf x}\cdot{\bf G}^{v,v}_{m,k}=-({\bf S}^{v,v}_k)^m,\eqn{24.2}$$
$$L^h({\bf G}^{v,v}_{m,k},{\bf G}^{v,h}_{m,k})=
-2\eta{\partial{\bf S}^{v,h}_k\over\partial{\bf x}_m}$$
$$-{\bf V}({\bf S}^{v,h}_k)^m+{\bf V}^m{\bf S}^{v,h}_k+{\bf H}({\bf S}^{v,v}_k)^m
-{\bf H}^m{\bf S}^{v,v}_k-{\bf H}^m{\bf e}_k;\eqn{24.3}$$
$$L^v({\bf G}^{h,v}_{m,k},{\bf G}^{h,h}_{m,k},G^{h,p}_{m,k})=
-2\nu{\partial{\bf S}^{h,v}_k\over\partial{\bf x}_m}+{\bf H}^k{\bf e}_m
-{\bf H}^m{\bf e}_k-({\bf V}\cdot{\bf S}^{h,v}_k){\bf e}_m$$
$$+{\bf V}^m{\bf S}^{h,v}_k+{\bf e}_mS^{h,p}_k
+({\bf H}\cdot{\bf S}^{h,h}_k){\bf e}_m-{\bf H}^m{\bf S}^{h,h}_k\eqn{25.1}$$
$$\nabla_{\bf x}\cdot{\bf G}^{h,v}_{m,k}=-({\bf S}^{h,v}_k)^m,\eqn{25.2}$$
$$L^h({\bf G}^{h,v}_{m,k},{\bf G}^{h,h}_{m,k})=
-2\eta{\partial{\bf S}^{h,h}_k\over\partial{\bf x}_m}$$
$$-{\bf V}({\bf S}^{h,h}_k)^m+{\bf V}^m{\bf S}^{h,h}_k+{\bf H}({\bf S}^{h,v}_k)^m
-{\bf H}^m{\bf S}^{h,v}_k+{\bf V}^m{\bf e}_k\eqn{25.3}$$

Взяв дивергенцию уравнений (24.3) и (25.3), и сравнив результат
с $m-$ми компонентами уравнений (11.3) и (12.3), соответственно, находим\break
$\nabla_{\bf x}\cdot{\bf G}^{v,h}_{m,k}=-({\bf S}^{v,h}_k)^m$
и $\nabla_{\bf x}\cdot{\bf G}^{h,h}_{m,k}=-({\bf S}^{h,h}_k)^m$.
Вследствие (21) и (22) правые части уравнений (24) и (25) являются
антисимметричными полями, поэтому выполнены следующие соотношения симметрии:
$${\bf G}^{v,v}_{m,k}({\bf x})=-{\bf G}^{v,v}_{m,k}(-{\bf x}),
\ \ {\bf G}^{v,h}_{m,k}({\bf x})=-{\bf G}^{v,h}_{m,k}(-{\bf x}),
\ \ G^{v,p}_{m,k}({\bf x})=G^{v,p}_{m,k}(-{\bf x}),$$
$${\bf G}^{h,v}_{m,k}({\bf x})=-{\bf G}^{h,v}_{m,k}(-{\bf x}),
\ \ {\bf G}^{h,h}_{m,k}({\bf x})=-{\bf G}^{h,h}_{m,k}(-{\bf x}),
\ \ G^{h,p}_{m,k}({\bf x})=G^{h,p}_{m,k}(-{\bf x}).$$

8. Средние части уравнений (17) при $n=2$ можно тогда
в силу (7), (6.2), (23.1) и (23.2) представить в виде
$$\nu\Delta_{\bf y}{\bf v}_0+\sum_{k=1}^3\sum_{m=1}^3\sum_{j=1}^3({\bf D}^{v,v}_{j,m,k}
{\partial^2{\bf v}^k_0\over\partial{\bf y}_m\partial{\bf y}_j}
+{\bf D}^{h,v}_{j,m,k}{\partial^2{\bf h}^k_0\over\partial{\bf y}_m\partial{\bf y}_j})
-\nabla_{\bf y}p'_1=\lambda_2{\bf v}_0,\eqn{26.1}$$
$$\eta\Delta_{\bf y}{\bf h}_0+\sum_{k=1}^3\sum_{m=1}^3
(\nabla_{\bf y}{\partial{\bf v}^k_0\over\partial{\bf y}_m}\times {\bf D}^{v,h}_{m,k}
+\nabla_{\bf y}{\partial{\bf h}^k_0\over\partial{\bf y}_m}\times {\bf D}^{h,h}_{m,k})
=\lambda_2{\bf h}_0.\eqn{26.2}$$
Здесь обозначено
$$p'_1=p_1-\sum_{k=1}^3\sum_{m=1}^3\left(\langle{\bf V}\cdot{\bf G}^{v,v}_{m,k}
-{\bf H}\cdot{\bf G}^{v,h}_{m,k}\rangle{\partial{\bf v}^k_0\over\partial{\bf y}_m}
+\langle{\bf V}\cdot{\bf G}^{h,v}_{m,k}-{\bf H}\cdot{\bf G}^{h,h}_{m,k}\rangle
{\partial{\bf h}^k_0\over\partial{\bf y}_m}\right),$$
$${\bf D}^{v,v}_{j,m,k}=\langle
-{\bf V}^j{\bf G}^{v,v}_{m,k}-{\bf V}({\bf G}^{v,v}_{m,k})^j
+{\bf H}^j{\bf G}^{v,h}_{m,k}+{\bf H}({\bf G}^{v,h}_{m,k})^j\rangle,$$
$${\bf D}^{h,v}_{j,m,k}=\langle
-{\bf V}^j{\bf G}^{h,v}_{m,k}-{\bf V}({\bf G}^{h,v}_{m,k})^j
+{\bf H}^j{\bf G}^{h,h}_{m,k}+{\bf H}({\bf G}^{h,h}_{m,k})^j\rangle,$$
$${\bf D}^{v,h}_{m,k}=\langle{\bf V}\times{\bf G}^{v,h}_{m,k}-{\bf H}\times{\bf G}^{v,v}_{m,k}\rangle,$$
$${\bf D}^{h,h}_{m,k}=\langle{\bf V}\times{\bf G}^{h,h}_{m,k}-{\bf H}\times{\bf G}^{h,v}_{m,k}\rangle.$$

Собственные функции задачи (26) -- гармоники Фурье:\break${\bf v}_0=\hat{\bf v}_0e^{i\xi\bf y}$
и ${\bf h}_0=\hat{\bf h}_0e^{i\xi\bf y}$,
где $\xi$ -- некоторый постоянный волновой вектор, а
векторы $\hat{\bf v}_0$ и $\hat{\bf h}_0$ удовлетворяют (15) и уравнениям
$$-\nu|\xi|^2\hat{\bf v}_0+\xi\times\left(\xi\times
\sum_{k=1}^3\sum_{m=1}^3\sum_{j=1}^3(\hat{\bf v}^k_0{\bf D}^{v,v}_{j,m,k}
+\hat{\bf h}^k_0{\bf D}^{h,v}_{j,m,k}){\xi_m\xi_j\over|\xi|^2}\right)=\lambda_2\hat{\bf v}_0,\eqn{27.1}$$
$$-\eta|\xi|^2\hat{\bf h}_0-\xi\times\sum_{k=1}^3\sum_{m=1}^3\xi_m
(\hat{\bf v}^k_0{\bf D}^{v,h}_{m,k}+\hat{\bf h}^k_0{\bf D}^{h,h}_{m,k})
=\lambda_2\hat{\bf h}_0.\eqn{27.2}$$

9. В дальнейшем будем предполагать для простоты, что $\lambda_2$ --
собственное значение задачи (15), (27) кратности 1 (это условие выполнено
для системы (1) общего положения).

Для решения уравнений (17) при $n>1$ сделаем подстановки
$${\bf w}_n={\bf w}''_n+\sum_{k=1}^3({\bf S}_k^{v,v}{\bf v}^k_n+{\bf S}^{h,v}_k{\bf h}^k_n)
+\sum_{k=1}^3\sum_{m=1}^3({\bf G}^{v,v}_{m,k}{\partial{\bf v}^k_{n-1}\over\partial{\bf y}_m}
+{\bf G}^{h,v}_{m,k}{\partial{\bf h}^k_{n-1}\over\partial{\bf y}_m}),\eqn{28.1}$$
$${\bf g}_n={\bf g}''_n+\sum_{k=1}^3({\bf S}_k^{v,h}{\bf v}^k_n+{\bf S}^{h,h}_k{\bf h}^k_n)
+\sum_{k=1}^3\sum_{m=1}^3({\bf G}^{v,h}_{m,k}{\partial{\bf v}^k_{n-1}\over\partial{\bf y}_m}
+{\bf G}^{h,h}_{m,k}{\partial{\bf h}^k_{n-1}\over\partial{\bf y}_m}),\eqn{28.2}$$
$$q_n=q''_n+\sum_{k=1}^3(S^{v,p}_k{\bf v}^k_n+S^{h,p}_k{\bf h}^k_n)
+\sum_{k=1}^3\sum_{m=1}^3(G^{v,p}_{m,k}{\partial{\bf v}^k_{n-1}\over\partial{\bf y}_m}
+G^{h,p}_{m,k}{\partial{\bf h}^k_{n-1}\over\partial{\bf y}_m}).\eqn{28.3}$$
Уравнения принимают вид
$$L^v({\bf w}''_n,{\bf g}''_n,q''_n)
+\nu\left(2(\nabla_{\bf x}\cdot\nabla_{\bf y}){\bf w}''_{n-1}
+\Delta_{\bf y}({\bf v}_{n-2}+{\bf w}_{n-2})\right.$$
$$\left.+2\sum_{k=1}^3\sum_{m=1}^3(
(\nabla_{\bf y}{\partial{\bf v}^k_{n-2}\over\partial{\bf y}_m}\cdot\nabla_{\bf x}){\bf G}^{v,v}_{m,k}
+(\nabla_{\bf y}{\partial{\bf h}^k_{n-2}\over\partial{\bf y}_m}\cdot\nabla_{\bf x}){\bf G}^{h,v}_{m,k})\right)$$
$$+{\bf V}\times\left(\sum_{k=1}^3\sum_{m=1}^3
(\nabla_{\bf y}{\partial{\bf v}^k_{n-2}\over\partial{\bf y}_m}\times{\bf G}^{v,v}_{m,k}
+\nabla_{\bf y}{\partial{\bf h}^k_{n-2}\over\partial{\bf y}_m}\times{\bf G}^{h,v}_{m,k})
+\nabla_{\bf y}\times{\bf w}''_{n-1}\right)$$
$$+\left(\sum_{k=1}^3\sum_{m=1}^3
(\nabla_{\bf y}{\partial{\bf v}^k_{n-2}\over\partial{\bf y}_m}\times{\bf G}^{v,h}_{m,k}
+\nabla_{\bf y}{\partial{\bf h}^k_{n-2}\over\partial{\bf y}_m}\times{\bf G}^{h,h}_{m,k})
+\nabla_{\bf y}\times{\bf g}''_{n-1}\right)\times{\bf H}$$
$$-\nabla_{\bf y}\left(p_{n-1}+q''_{n-1}
+\sum_{k=1}^3\sum_{m=1}^3({\partial{\bf v}^k_{n-2}\over\partial{\bf y}_m}G^{v,p}_{m,k}
+{\partial{\bf h}^k_{n-2}\over\partial{\bf y}_m}G^{h,p}_{m,k})\right)$$
$$-\sum_{m=0}^{n-2}\lambda_{n-m}({\bf v}_m+{\bf w}_m)=0,\eqn{29.1}$$
$$L^h({\bf w}_n,{\bf g}_n)
+\eta\left(2(\nabla_{\bf x}\cdot\nabla_{\bf y}){\bf g}''_{n-1}
+\Delta_{\bf y}({\bf h}_{n-2}+{\bf g}_{n-2})\right.$$
$$\left.+2\sum_{k=1}^3\sum_{m=1}^3(
(\nabla_{\bf y}{\partial{\bf v}^k_{n-2}\over\partial{\bf y}_m}\cdot\nabla_{\bf x}){\bf G}^{v,h}_{m,k}
+(\nabla_{\bf y}{\partial{\bf h}^k_{n-2}\over\partial{\bf y}_m}\cdot\nabla_{\bf x}){\bf G}^{h,h}_{m,k})\right)$$
$$+\sum_{k=1}^3\sum_{m=1}^3\left(
\nabla_{\bf y}{\partial{\bf v}^k_{n-2}\over\partial{\bf y}_m}\times
({\bf V}\times{\bf G}^{v,h}_{m,k}+{\bf G}^{v,v}_{m,k}\times{\bf H})\right.$$
$$\left.+\nabla_{\bf y}{\partial{\bf h}^k_{n-2}\over\partial{\bf y}_m}\times
({\bf V}\times{\bf G}^{h,h}_{m,k}+{\bf G}^{h,v}_{m,k}\times{\bf H})\right)$$
$$+\nabla_{\bf y}\times({\bf V}\times{\bf g}''_{n-1}+{\bf w}''_{n-1}\times{\bf H})
-\sum_{m=0}^{n-2}\lambda_{n-m}({\bf h}_m+{\bf g}_m)=0.\eqn{29.2}$$
Покажем, что можно последовательно найти из них все
компоненты асимптоти\-ческих разложений (3)-(5), причем выполнены
соотношения (18). Пусть для неко\-торого $N$ решены все уравнения при $n<N$, в результате
этого определены ${\bf v}_n,{\bf w}_n,{\bf h}_n,{\bf g}_n$ и $q_n$ при всех $n<N-2$,
$p_n$ при всех $n<N-1$, а также ${\bf w}''_n,{\bf g}''_n,q''_n$ и
$\lambda_n$ при всех $n<N$, и зависимость всех найденных неизвестных функций
от медленной переменной выражается в их пропорциональности $e^{i\xi\bf y}$.
Рассмотрим уравнение, соответствующее $n=N$.

Средние части уравнений (29) при $n=N$ имеют вид
$$\nu\Delta_{\bf y}{\bf v}_{N-2}+\sum_{k=1}^3\sum_{m=1}^3\sum_{j=1}^3({\bf D}^{v,v}_{j,m,k}
{\partial^2{\bf v}^k_{N-2}\over\partial{\bf y}_m\partial{\bf y}_j}
+{\bf D}^{h,v}_{j,m,k}{\partial^2{\bf h}^k_{N-2}\over\partial{\bf y}_m\partial{\bf y}_j})$$
$$-\nabla_{\bf y}p_{N-1}-\lambda_2{\bf v}_{N-2}-\lambda_N{\bf v}_0$$
$$=-\langle{\bf V}\times(\nabla_{\bf y}\times{\bf w}''_{N-1})
+(\nabla_{\bf y}\times{\bf g}''_{N-1})\times{\bf H}\rangle
+\sum_{m=1}^{N-3}\lambda_{N-m}{\bf v}_m,\eqn{30.1}$$
$$\eta\Delta_{\bf y}{\bf h}_{N-2}+\sum_{k=1}^3\sum_{m=1}^3
(\nabla_{\bf y}{\partial{\bf v}^k_{N-2}\over\partial{\bf y}_m}\times{\bf D}^{v,h}_{m,k}
+\nabla_{\bf y}{\partial{\bf h}^k_{N-2}\over\partial{\bf y}_m}\times{\bf D}^{h,h}_{m,k})
-\lambda_2{\bf h}_{N-2}-\lambda_N{\bf h}_0$$
$$=-\nabla_{\bf y}\times\langle{\bf V}\times{\bf g}''_{N-1}+{\bf w}''_{N-1}\times{\bf H}\rangle
+\sum_{m=1}^{N-3}\lambda_{N-m}{\bf h}_m.\eqn{30.2}$$
Согласно предположению индукции правые части (30) имеют вид
$\hat{\bf f}^v_Ne^{i\xi\bf y}$ и $\hat{\bf f}^h_Ne^{i\xi\bf y}$, где
векторы-константы $\hat{\bf f}^v_N$ и $\hat{\bf f}^h_N$ известны, причем
$\hat{\bf f}^h_N\cdot\xi=0$. Обозначим матрицу размера $6\times6$ в левой части
(27) через $D(\xi)$. Подставляя (18.1) при $n=N-2$ и сокращая
множитель $e^{i\xi\bf y}$, приводим (30) к эквивалентной системе
$$(D(\xi)-\lambda_2)\left(\hat{\bf v}_{N-2}\atop\hat{\bf h}_{N-2}\right)
-\lambda_N\left(\hat{\bf v}_0\atop\hat{\bf h}_0\right)
=\left(-|\xi|^{-2}\xi\times(\xi\times\hat{\bf f}^v_N)
\atop\hat{\bf f}^h_N\right),$$
которую решаем в $V_\xi$ аналогично задаче (20).
Именно, находим $\lambda_N$ и вектор
$(\hat{\bf v}_{N-2},\hat{\bf h}_{N-2})$ в собственном подпространстве
пространства $V_\xi$, дополнительном к подпространству, натянутому на
$(\hat{\bf v}_0,\hat{\bf h}_0)$. Затем из (30.1) находим $p_{N-1}$,
и из (28) при $n=N-2$ -- функции ${\bf w}_{N-2},{\bf g}_{N-2}$ и $q_{N-2}$.

Осциллирующие части уравнений (29) при $n=N$
рассматриваем при условиях, следующих из (6.2):
$$\nabla_{\bf x}\cdot{\bf w}''_N=-\nabla_{\bf y}\cdot{\bf w}''_{N-1}
-\sum_{k=1}^3\sum_{m=1}^3({\bf G}^{v,v}_{m,k}\cdot\nabla_{\bf y}{\partial{\bf v}^k_{N-2}\over\partial{\bf y}_m}
+{\bf G}^{h,v}_{m,k}\cdot\nabla_{\bf y}{\partial{\bf h}^k_{N-2}\over\partial{\bf y}_m}),$$
$$\nabla_{\bf x}\cdot{\bf g}''_N=-\nabla_{\bf y}\cdot{\bf g}''_{N-1}
-\sum_{k=1}^3\sum_{m=1}^3({\bf G}^{v,h}_{m,k}\cdot\nabla_{\bf y}{\partial{\bf v}^k_{N-2}\over\partial{\bf y}_m}
+{\bf G}^{h,h}_{m,k}\cdot\nabla_{\bf y}{\partial{\bf h}^k_{N-2}\over\partial{\bf y}_m}),$$
позволяющих выделить градиентные части неизвестных вектор-функций
${\bf w}''_N$ и ${\bf g}''_N$ и свести данную задачу к задаче вида (8)
для соленоидальных неизвестных.

Таким образом, построено полное асимптотическое разложение мод линейных
возмущений и их временных инкрементов для случая, когда невозмущенное исходное
стационарное состояние имеет центр симметрии. Если предельный опе\-ратор
$L_\infty$, определенный левой частью (26), эллиптичен, то можно доказать
аналогично [Вишик, 1986], что
в указанном случае каждому собственному значе\-нию оператора $L_\infty$
отвечает ветвь собственных значений исходного оператора (задачи (2)).

Анализируя приведенные алгебраические построения, можно показать (мето\-дом
математической индукции), что $(i)$ в разложениях (3) все члены при четных
степенях $\epsilon$ симметричны, а все члены при нечетных степенях
антисимметричны (и следовательно ${\bf v}_{2n+1}={\bf h}_{2n+1}=0$ при
всех целых $n$); $(ii)$ в разложении (4) все члены при четных степенях
$\epsilon$ антисимметричны (и следовательно $p_{2n}=0$ при всех $n$),
а все члены при нечетных степенях симметричны; $(iii)$ $\lambda_{2n+1}=0$
при всех целых $n$, т.е. собственные значения $\lambda$ являются рядами
по степеням $\epsilon^2$.

10. При решении задачи об устойчивости для некоторого заданного стацио\-нарного
МГД состояния ${\bf V},{\bf H},P$ решения вспомогательных задач (11), (12),
(24) и (25), определяющие коэффициенты тензора диффузии $\bf D$, можно найти
числен\-но. Величину $\min_{|\xi|=1}(-\lambda_2(\xi))$ можно
интерпретировать, как минимальный коэф\-фициент вихревой диффузии.

Аналогично работе [Zheligovsky, Podvigina, Frisch, 2001] был исследован вопрос
о том, насколько часто эффект отрицательной вихревой диффузии воспроизво\-дится
в модельных случаях. Стационарные поля $\bf V,H$ задавались в виде рядов Фурье
со случайными гармониками (спроектированных на пространство
соленои\-дальных вектор-функций) и заданным энергетическим спектром. Энергия
гар\-моник экспо\-ненциально убывала на 6 порядков от 1-го до 10-го сферического
слоя в пространстве гармоник, ряд обрывался на волновых векторах длины 10,
а среднеквадратичная величина энергии поля была равна 1.
Решения вспомогатель\-ных задач искали итерационными методами [Желиговский, 2001]
в виде рядов Фурье с разрешением $64^3$ гармоник, что обеспечивало достаточную
точность (затухание спектра решений составляло 18-20 порядков).

Гистограммы полученных значений минимального коэффициента вихревой диффузии
приведены на Рис.~1. Эффект отрицательной вихревой
диффузии наблюдается уже при \hbox{$\nu=\eta=3/4$}, что соответствует мелкомасштабным
кинема\-тическим и магнитным числам Рейнольдса $4/3$ (при их определении
характерный про\-странственный масштаб взят равным 1, порядка размера куба
периодичности стационарного состояния). Это интересно сравнить
с результатами [Zheligovsky, Podvigina, Frisch, 2001], согласно которым
эффект отрицательной вихревой магнитной диффузии в задаче кинематического
магнитного динамо возникает при существенно более низких значениях коэффициента
молекулярной магнитной диффузии: порог $\eta$, при которых вихревая магнитная
диффузия может стать отрицательной, лежит в интервале $0.2<\eta<0.3$ .

\pagebreak
\centerline{\psfig{file=figaxx.ps,height=70mm}\hspace{5mm}\psfig{file=figbxx.ps,height=70mm}}

\centerline{(a)\hspace{75mm}(b)}

~

\centerline{\psfig{file=figcxx.ps,height=70mm}}

\centerline{(c)}

~

\noindent
Рис. 1. Гистограммы значений минимального коэффициента вихревой диффузии
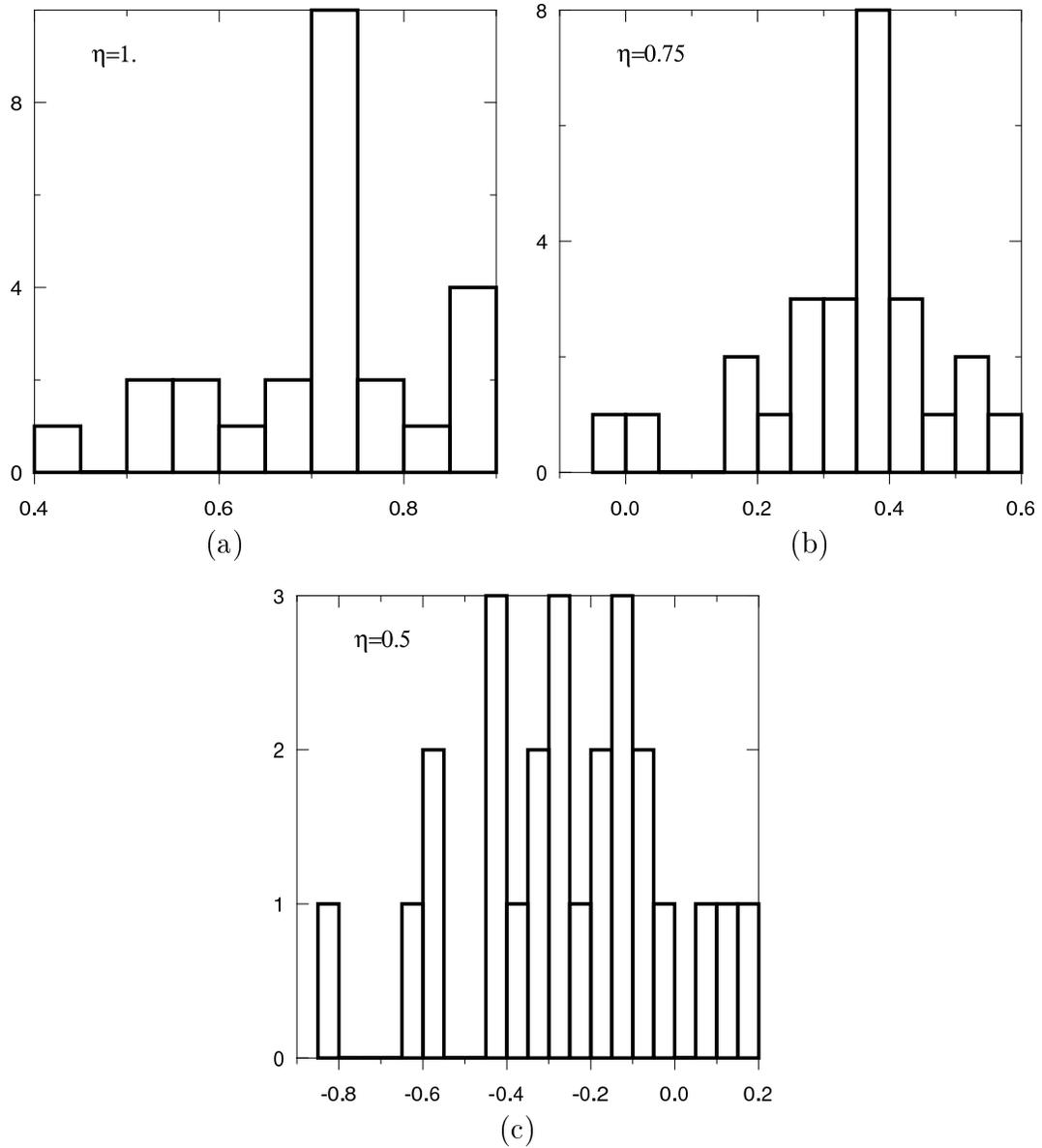
в ансамблях из 25 стационарных МГД состояний со случайными гармониками:
$\nu=\eta=1$ (a), $\nu=\eta=3/4$ (b), $\nu=\eta=1/2$ (c).

\mi{\bf Благодарности}.
Работа выполнена с использованием вычислительных средств, предоставленных
французской программой ``Simulations Interactives et Visualisation
en Astronomie et M\'ecanique (SIVAM)" в Обсерватории Лазурного Берега (Ницца).
Визит автора финансировался Министерством образования Франции.

\pagebreak
\mi{\bf Литература}

\mi {\it Вишик M.M.} Периодическое динамо. I // Математические методы
сейсмологии и геодинамики (Вычислительная Сейсмология,
вып. 19). М.: Наука, 186-215 (1986).

\mi {\it Вишик M.M.} Периодическое динамо. II // Численное моделирование
и анализ геофизических процессов (Вычислительная Сейсмология.
вып. 20). М.: Наука, 12-22 (1987).

\mi {\it Желиговский В.А.} О генерации магнитного поля движением проводящей
среды, имеющим внутренний масштаб // Компьютерный анализ геофизических полей
(Вычислительная сейсмология, вып. 23). М.: Наука, 161-181 (1990).

\mi {\it Желиговский В.А.} О генерации магнитного поля движением проводящей
среды, имеющим внутренний масштаб. II // Современные методы обработки
сейсмоло\-гических данных (Вычислительная сейсмология, вып. 24).
М.: Наука, 205-217 (1991).

\mi {\it Желиговский В.А.} Чебышевский итерационный метод с расщеплением
оператора для вычисления корней больших систем уравнений //
Труды международной конференции "Нелинейные задачи теории гидродинамической
устойчивости и турбулентности", Москва, 13-17 февраля 2000.
под ред. С.Я.Герценштейн, Изд-во МГУ, 2001
(см. http://www.obs-nice.fr/etc7/vlad.ps.gz).

\mi {\it Жиков В.В., Козлов С.М., Олейник O.A., Ха Тьен Нгоан.}
Усреднение и $G$-сходимость дифференциальных операторов //
УМН. 34 $N5$, 63-133 (1979).

\mi {\it Като Т.} Теория возмущения линейных операторов. М.: Мир (1972).

\mi {\it Козлов С.М.} Усреднение дифференциальных операторов с почти
периодическими быстро осциллирующими коэффициентами // Мат. сборник.
107 $N2$, 199-217 (1978).

\mi {\it Bensoussan A., Lions J.-L., Papanicolaou G.} Asymptotic
Analysis for Periodic Structures. North Holland (1978).

\mi {\it Biferale L., Crisanti A., Vergassola M., Vulpiani A.} Eddy viscosity
in scalar transport // Phys. Fluids. 7 $N11$, 2725-2734 (1995).

\mi {\it Chandrasekhar S.} Hydrodynamic and hydromagnetic stability.
Oxford Univ. Press (1961).

\mi {\it Dubrulle B., Frisch U.} Eddy viscosity of parity-invariant flow //
Phys. Rev. A. 43 $N10$, 5355-5364 (1991).

\mi {\it Frisch U., Zhen Su She, Sulem P.L.} Large-scale flow driven by the
anisotropic kinetic alpha effect // Physica D. 28, 382-392 (1987).

\mi {\it Gama S., Vergassola M., Frisch U.} Negative eddy viscosity in
isotropically forced two-dimensional flow: linear and nonlinear dynamics //
J. Fluid Mech. 260, 95-126 (1994).

\mi {\it Glatzmaier G.A., Roberts P.H.} A three-dimensional self-consistent
computer simulation of a geomagnetic field reversal //
Nature. 377, 203-209 (1995).

\mi {\it Glatzmaier G.A., Roberts P.H.} An anelastic geodynamo simulation
driven by compositional and thermal convection // Physica D. 97, 81-94 (1996).

\mi {\it Lanotte A., Noullez A., Vergassola M., Wirth A.}
Large-scale dynamo by negative magnetic eddy diffusivities //
Geophys. Astrophys. Fluid Dynamics. 91, 131-146 (1999).

\mi {\it Roberts P.H., Glatzmaier G.A.} The geodynamo, past, present
and future // Geophys.~Astrophys.~Fluid Dynamics. 94, 47--84 (2001).

\mi {\it Sarson G.R., Jones C.A.} A convection driven geodynamo reversal
model // Phys.~Earth Planet.~Int. 111, 3-20 (1999).

\mi {\it Sivashinsky G., Frenkel A.} Negative eddy viscosity under conditions
of isotropy // Phys. Fluids. A4 $N8$, 1608-1610 (1992).

\mi {\it Sivashinsky G., Yakhot V.} Negative viscosity effect in large-scale
flows // Phys. Fluids. 28 $N4$, 1040-1042 (1985).

\mi {\it Sulem P.L., She Z.S., Scholl H., Frisch U.} Generation of
large-scale structures in three-dimensional flow lacking parity-invariace //
J. Fluid Mech. 205, 341-358 (1989).

\mi {\it Vergassola M., Avellaneda M.} Scalar transport in compressible
flow // Physica D. 106, 148-166 (1997).

\mi {\it Wirth A., Gama S., Frisch U.} Eddy viscosity of
three-dimensional flow // J. Fluid Mech. 288, 249-264 (1995).

\mi {\it Zheligovsky V.A.} $\alpha$-effect in generation of magnetic field
by a flow of conducting fluid with internal scale in an axisymmetric volume //
Geophys. Astrophys. Fluid Dynamics. 59, 235-251 (1991).

\mi {\it Zheligovsky V.A., Podvigina O.M., Frisch U.} Dynamo effect
in parity-invariant flow with large and moderate separation of scales //
Geophys. Astrophys. Fluid Dynamics. 95, 227-268 (2001)
[http://xxx.lanl.gov/abs/nlin.CD/0012005].
\end{document}